\documentclass[11pt,twoside]{article}
\usepackage{mathrsfs}
\usepackage{amsmath}
\usepackage{amssymb}
\usepackage{graphicx}
\usepackage{subfigure}
\usepackage{appendix}
\usepackage{cite}
\numberwithin{equation}{section}

\newtheorem{proposition}{Proposition}
\newtheorem{lemma}{Lemma}

\newtheorem{corollary}{Corollary}

\DeclareMathOperator{\im}{Im}
\DeclareMathOperator{\re}{Re}

\DeclareMathOperator{\diag}{diag}
\newcommand*{\QEDB}{\hfill\ensuremath{\square}}

\title{Action-angle variables for the nonlinear Schr\"{o}dinger equation on the half-line}
\author{
Baoqiang Xia
\\
School of Mathematics and Statistics, Jiangsu Normal University,
\\
Xuzhou, Jiangsu 221116, P. R. China,
\\
E-mail address: xiabaoqiang@126.com
}

\date{}
\begin{document}
\maketitle
\begin{abstract}
We consider the nonlinear Schr\"{o}dinger (NLS) equation on the half-line subjecting to a class of boundary conditions preserve the integrability of the model. For such a half-line problem, the Poisson brackets of the corresponding scattering data are computed, and the variables of action-angle type are constructed. These action-angle variables completely trivialize the dynamics of the NLS equation on the half-line.


%\noindent{\bf PACS numbers:}\quad 02.30.Ik, 02.30.Jr
\end{abstract}
\newpage

\section{ Introduction}

Integrable nonlinear partial differential equations (PDEs) in $1+1$-dimensions serve as infinite-dimensional analogues of Hamiltonian systems in classical mechanics.
To investigate the Liouville integrability of such PDEs in the presence of boundaries, suitable boundary conditions must be imposed. For $1+1$-dimensional PDEs with the space variable posed on the half-line or the finite interval, an important step was taken by Sklyanin in \cite{Sklyanin1987} to introduce a systematic approach to selecting boundary conditions that preserve the integrability of a model. In Sklyanin's formalism, the admissible boundary conditions are described by solutions of the classical reflection equations: the reflection matrices \cite{Sklyanin1987} (see also \cite{ACC2018} for a recent development). For the boundary models constructed in this manner, the integrability is ensured by the existence of an infinite set of integrals of the motion in involution.

In addition to the integrability aspect, another important topic is the solution method to the integrable boundary problem.
%The Sklyanin's approach works well on integrability aspect (address the integrability aspect clearly), but did not tackle the solution method.
In this respect, several methods have been introduced to solve the integrable boundary value problems defined on the half-line or the finite interval, such as the nonlinear mirror image method developed e.g. in \cite{BH2009,BB2012,CZ2012,CCD2021,CC2019} following the idea initiated in \cite{Habibullin1991,BT1989,BT1991,T1991}, a boundary dressing technique presented recently in \cite{Zhang2019}, and a more recent method of \cite{Zhangc2021,CCRZ2022} which is based on Sklyanin's double-row monodromy matrix.
The first and third methods in the aforementioned ones extend the applications of the well-known inverse scattering transform (IST) (see e.g. \cite{GGKM1967,ZS1972,AKNS1,AKNS2}) from the full-line case to the half-line case with integrable boundary conditions.
We note that a more general method to analyze boundary value problems, known as unified transform method, was due to Fokas \cite{Fokas}. The effectiveness of Fokas' unified transform method is manifested when one is working with boundary conditions which break integrability at the boundary.

For the full-line problem with vanishing boundaries, an important discovery is that the IST provides an infinite-dimensional analogue of Liouville theorem in classical mechanics:
it provides the variables of action-angle type, which linearize the model, in the infinite-dimensional Hamiltonian setting, see e.g. \cite{ZF1971,ZM1974,BFT1986,Faddeev2007,CI2006}.
The main purpose of the present paper is to carry out this programme to the half-line problem in the presence of integrable boundary conditions. The particular model we consider is the NLS equation
\begin{eqnarray}
iu_t+u_{xx}+2 u|u|^2=0,
\label{nls}
\end{eqnarray}
%subjecting to a class of integrable boundary conditions constructed from Sklyanin's approach.
where $u\equiv u(x,t)$ is a complex-valued function and $|u|^2=u\bar{u}$ with the bar denoting complex conjugation.
We concentrate on a class of integrable boundary conditions introduced in \cite{Zhangc2021,CCRZ2022}, which include the well-known Robin boundary condition \cite{Sklyanin1987} and a new time-dependent boundary condition \cite{Zambon2014} attracted attention recently \cite{Xia2021,CCD2021,CCRZ2022,Gruner2020}, as special cases.
For the NLS equation equipped with such boundary conditions, we compute the Poisson brackets of the corresponding scattering data, and we construct explicit variables of action-angle type which trivialize the dynamics of the NLS equation on the half-line.

The paper is arranged as follows.
In section 2, we collect some basic results regarding the Hamiltonian formulation of the NLS equation on the full-line that we will need. In section 3, we concentrate on the NLS equation posed on the half-line with a class of integrable boundary conditions and investigate the corresponding inverse scattering transform. Section 4 concerns our main results: we compute the Poisson brackets between all the elements of the scattering data and derive the variables of action-angle type for the half-line problem discussed in section 3. We discuss our results further in section 5.

\section{The NLS equation on the full-line}

We start by briefly outline the main results from the Hamiltonian formulation of the NLS equation on the full $x$-axis under the assumption that the field and its derivatives decrease rapidly as $|x|\rightarrow\infty$.

The NLS equation admits a Lax pair \cite{AKNS1,AKNS2}
\begin{subequations}
\begin{eqnarray}
\phi_x(x,t,\lambda)=U(x,t,\lambda)\phi(x,t,\lambda),
~~
U(x,t,\lambda)=\left( \begin{array}{cc} \frac{\lambda}{2i} & -\bar{u} \\
 u &  -\frac{\lambda}{2i} \\ \end{array} \right),
 \label{lpx}
\\
\phi_t(x,t,\lambda)=V(x,t,\lambda)\phi(x,t,\lambda),
~~ V(x,t,\lambda)=\left( \begin{array}{cc}  \frac{i}{2}\lambda^2-i|u|^2 & \lambda \bar{u}+i\bar{u}_x \\
 -\lambda u+iu_x & -\frac{i}{2}\lambda^2+i|u|^2  \\ \end{array} \right),
 \label{lpt}
\end{eqnarray}
\label{lpxt}
\end{subequations}
where $\lambda$ is a spectral parameter.

An important quantity, called the transition matrix \cite{Faddeev2007}
\begin{eqnarray}
T(x,y,\lambda)=\overset{\curvearrowleft}\exp \int_{y}^{x} U(\xi,\lambda)d\xi,
\end{eqnarray}
is defined as the fundamental solution of the space-part of the Lax equations
\begin{eqnarray}
\frac{\partial T(x,y,\lambda)}{\partial x}=U(x,\lambda)T(x,y,\lambda),
\label{Tx}
\end{eqnarray}
with the initial condition
\begin{eqnarray}
\left.T(x,y,\lambda)\right|_{x=y}=I.
\end{eqnarray}
The transition matrix $T(x,y,\lambda)$ has an integral representation \cite{Faddeev2007},
\begin{eqnarray}
T(x,y,\lambda)=E(x-y,\lambda)+\int_{y}^{2x-y}E(x-z,\lambda)\mathcal{T}(x,y,z)dz,
\label{Tir}
\end{eqnarray}
where
\begin{eqnarray}
E(x,\lambda)=\exp\left\{\frac{\lambda}{2i}x\sigma_3\right\}, ~~\sigma_3=\diag(1,-1),
\end{eqnarray}
and the matrix valued kernel $\mathcal{T}$ satisfies
\begin{eqnarray}
\mathcal{T}(x,y,z)=\frac{1}{2}U_0(\frac{y+z}{2})+\int_{\frac{y+z}{2}}^{x}U_0(s)\mathcal{T}(s,y,2s-z)ds,
\label{Tker}
\end{eqnarray}
with $U_0(x)\equiv U(x,\lambda)+\frac{i\lambda}{2}\sigma_3$.
The evolution of $T(x,y,\lambda)$ along $t$ is given by
\begin{eqnarray}
 \frac{\partial T(x,y,\lambda)}{\partial t}=V(x,\lambda)T(x,y,\lambda)
 -T(x,y,\lambda)V(y,\lambda).
 \label{Td}
\end{eqnarray}
The structure of $U$ implies that the transition matrix $T(x,y,\lambda)$ is unimodular
\begin{eqnarray}
\det T(x,y,\lambda)=1,
\label{detT}
\end{eqnarray}
and it satisfies the following involution relation
\begin{eqnarray}
T(x,y,\lambda)=\sigma \overline{T(x,y,\bar{\lambda})}\sigma,
~~\sigma=\left( \begin{array}{cc} 0 & -i \\
 i & 0  \\ \end{array} \right).
\label{Tinvol}
\end{eqnarray}
Indeed, the unimodular property (\ref{detT}) follows from the fact that $U$ is traceless, while the involution relation (\ref{Tinvol}) follows from the fact that $U$ satisfies $U(x,\lambda)=\sigma \overline{U(x,\bar{\lambda})}\sigma$.

Consider the canonical Poisson brackets \cite{Faddeev2007}
\begin{eqnarray}
\left\{u(x,t),u(y,t)\right\}=\left\{\bar{u}(x,t),\bar{u}(y,t)\right\}=0,
~~
\left\{u(x,t),\bar{u}(y,t)\right\}=i\delta(x-y),
\label{cPB}
\end{eqnarray}
where $\delta(x-y)$ is the Dirac $\delta$-function.
Using these Poisson brackets, one can deduce that the transition matrix satisfies the following well-known relation (see e.g. \cite{Faddeev2007})
\begin{eqnarray}
\left\{T_1(x,y,\lambda),T_2(x,y,\mu)\right\}=\left[r(\lambda-\mu),T_1(x,y,\lambda)T_2(x,y,\mu)\right],
\label{rmrelation1}
\end{eqnarray}
for $y<x$, where $T_1(x,y,\lambda)=T(x,y,\lambda)\otimes I$, $T_2(x,y,\mu)=I\otimes T(x,y,\mu)$, and the classical $r$-matrix $r$ is
\begin{eqnarray}
r(\lambda)=\frac{1}{\lambda}\left( \begin{array}{cccc}
1 & 0 & 0 & 0
\\
0 &  0 & 1 & 0
\\
0 & 1 & 0 & 0
\\
0 &  0 & 0 & 1
 \\ \end{array} \right).
 \label{rm}
\end{eqnarray}
For $x<y$, it follows from (\ref{rmrelation1}) that
\begin{eqnarray}
\left\{T_1(x,y,\lambda),T_2(x,y,\mu)\right\}=-\left[r(\lambda-\mu),T_1(x,y,\lambda)T_2(x,y,\mu)\right],
\label{rmrelation2}
\end{eqnarray}
since $T(y,x,\lambda)=T^{-1}(x,y,\lambda)$.

With the Poisson brackets (\ref{cPB}), the NLS equation on the full $x$-axis can be written in the Hamilton form
\begin{eqnarray}
u_t=\left\{H,u\right\},~~\bar{u}_t=\left\{H,\bar{u}\right\},
\label{HE}
\end{eqnarray}
with the Hamiltonian
\begin{eqnarray}
H=-\int_{-\infty}^{\infty}\left(|u|^4-|u_x|^2\right)dx.
\label{H}
\end{eqnarray}
Local integrals of the motion of the NLS equation with vanishing boundaries can be constructed from the monodromy matrix $T(\lambda)$ which is defined by \cite{Faddeev2007}
\begin{eqnarray}
T(\lambda)=\lim_{x\rightarrow \infty}\lim_{y\rightarrow -\infty}E(-x,\lambda)T(x,y,\lambda)E(y,\lambda).
\label{Tlambda}
\end{eqnarray}
Indeed, the quantity $\ln T_{11}(\lambda)$, where $T_{11}(\lambda)$ stands for the $11$-entry of $T(\lambda)$, provides a generating function for the local integrals of the motion. By investigating the large $\lambda$ expansion of $\ln T_{11}(\lambda)$, we can extract explicit forms of the local integrals of the motion order by order.
Moreover, with the $r$-matrix relation (\ref{rmrelation1}) or (\ref{rmrelation2}), one can prove that such integrals of the motion are in involution (see \cite{Faddeev2007} for details).
Thus, the integrability of the NLS equation with vanishing boundaries from the Hamiltonian standpoint is obtained.

\section{The NLS equation on the half-line}

We now proceed to the NLS equation on the positive $x$-axis subjecting to suitable boundary conditions at $x=0$ such that the integrability of the model is preserved. We concentrate on the case where the NLS field is of Schwartz type in $x$, i.e. $u(x,t_0)\in \mathcal{S}(\mathbb{R}^{+})$ for a fixed time $t=t_0$. This means that vanishing boundary conditions are imposed for the NLS field and its derivatives as $x\rightarrow\infty$.

%We consider the NLS equation on the positive $x$-axis subjecting to the boundary condition at $x=0$ as stated in section 3, under the assumption that the NLS field is of Schwartz type in $x$, i.e. $u(x,t_0)\in \mathcal{S}(\mathbb{R}^{+})$ for a fixed time $t=t_0$. This means that the NLS field is infinitely differentiable and together with all its derivatives vanishes as $x\rightarrow\infty$.

\subsection{Boundary conditions}

Following \cite{Sklyanin1987} (see also \cite{ACC2018}), we consider the boundary conditions at $x=0$ that are characterized by the matrices $K(\lambda)$ obeying both
\begin{eqnarray}
 \frac{dK(\lambda)}{dt}=V(0,t,\lambda)K(\lambda)-K(\lambda)V(0,t,-\lambda),
 \label{algc}
\end{eqnarray}
and
\begin{eqnarray}
\left\{K_1(\lambda),K_2(\mu)\right\}=\left[r(\lambda-\mu),K_1(\lambda)K_2(\mu)\right]
+K_1(\lambda)r(\lambda+\mu)K_2(\mu)-K_2(\mu)r(\lambda+\mu)K_1(\lambda),
 \label{alga}
\end{eqnarray}
where we assume that the reflection matrices $K(\lambda)$ can depend on time in general.
For a boundary condition resulting from (\ref{algc}) with $K$ obeying (\ref{alga}), one can establish the integrability of the corresponding half-line problem in the sense of the existence of infinitely many Poisson commuting conserved quantities by investigating the Sklyanin's double-row monodromy matrix. We refer the reader to \cite{Sklyanin1987,ACC2018} for details on this issue. In this paper, we will confirm further the integrability of such boundary problems by investigating the corresponding scattering problem (see section 3.3 below).

In the rest of the paper, we will require that $K(\lambda)$ satisfies the following normalization conditions
\begin{eqnarray}
K(\lambda)K(-\lambda)=I,~~\det\left(K(\lambda)\right) K(-\lambda)=\sigma \overline{K(-\bar{\lambda})}\sigma.
\label{Knorm}
\end{eqnarray}
Let
\begin{eqnarray}
d(\lambda)=\det K(-\lambda).
\label{dlambda}
\end{eqnarray}
From (\ref{Knorm}), we immediately obtain
\begin{eqnarray}
d(\lambda)d(-\lambda)=1,~~d(-\lambda)=\overline{d(\bar{\lambda})}.
\label{dsym}
\end{eqnarray}
We assume that $d(\lambda)$ is a rational function of $\lambda$. Then the normalization condition (\ref{dsym}) implies that $d(\lambda)$ takes the form
\begin{eqnarray}
d(\lambda)=\varepsilon\prod_{j=1}^N\frac{\left(\lambda-\beta_j\right)\left(\lambda+\bar{\beta}_j\right)}
{\left(\lambda-\bar{\beta}_j\right)\left(\lambda+\beta_j\right)}
\equiv\varepsilon\frac{f(\lambda)}{g(\lambda)},~~\varepsilon=\pm 1,
\label{d}
\end{eqnarray}
where
\begin{eqnarray}
f(\lambda)=g(-\lambda)=\prod_{j=1}^N\left(\lambda-\beta_j\right)\left(\lambda+\bar{\beta}_j\right).
\label{f}
\end{eqnarray}
We further require that the $K(\lambda)$ matrix subjects to the following form
\begin{eqnarray}
K(\lambda)=\frac{1}{f(\lambda)}\mathcal{K}(\lambda),
\label{K}
\end{eqnarray}
where the entries of $\mathcal{K}(\lambda)$ are polynomial of $\lambda$,
and satisfy the asymptotic
\begin{eqnarray}
\lim_{\lambda\rightarrow\infty} K(\lambda)=\diag(1,\varepsilon).
\label{Kasy}
\end{eqnarray}
The above normalization conditions will be needed when investigating the analytic properties of Jost solutions and scattering data for the corresponding half-line problem (see section 3.2 below).

A class of $K(\lambda)$ matrices meeting equation (\ref{algc}) was presented in \cite{CCRZ2022}. Here we only present the following interesting boundary conditions as examples.

{\bf Example 1: Robin boundary condition}.
Taking
\begin{eqnarray}
K(\lambda)=\left( \begin{array}{cc} 1 & 0 \\
 0 &  -\frac{\lambda+i\beta}{\lambda-i\beta} \\ \end{array} \right)
 \label{K1}
\end{eqnarray}
where $\beta$ is a real parameter,
we obtain from (\ref{algc}) the well-known Robin boundary condition (see e.g. \cite{Sklyanin1987,Habibullin1991})
\begin{eqnarray}
\left.\left(u_x+\beta u\right)\right|_{x=0}=0,
\label{rbc}
\end{eqnarray}
for the classical NLS equation (\ref{nls}). In this case, the Poisson bracket (\ref{alga}) automatically holds.
%The Robin boundary has been studied extensively in the literatures, see e.g. \cite{}.

{\bf Example 2: A boundary condition involving the time derivative of the field}.
Taking
\begin{eqnarray}
K(\lambda)=\left.\frac{1}{\lambda-i\beta}\left(\lambda\mathbf{I}+\left( \begin{array}{cc} -i\Omega & 2i\bar{u}
\\
 2i u &  i\Omega  \\ \end{array} \right)\right)\right|_{x=0},
 \label{K2}
\end{eqnarray}
where $\beta$ is a real parameter and
\begin{eqnarray}
\Omega=\sqrt{\beta^2 - 4|u|^2},
\label{Omg}
\end{eqnarray}
then we obtain from (\ref{algc}) the following boundary condition
\begin{eqnarray}
\left.\left(u_t-2i|u|^2u+i\Omega u_x\right)\right|_{x=0}=0,
\label{tbcnls}
\end{eqnarray}
which appeared in \cite{Zhangc2021,Xia2023}.
In this situation, the Poisson bracket (\ref{alga}) is not trivial, it is equivalent to the following boundary Poisson bracket
\begin{eqnarray}
\left\{u,\bar{u}\right\}=-i\Omega,
\label{BP1}
\end{eqnarray}
at $x=0$.

{\bf Example 3: Another boundary condition involving the time derivative of the field}.
Taking
\begin{eqnarray}
K(\lambda)=\left.\frac{1}{\left(\lambda+\alpha+i\beta\right)\left(\lambda-\alpha+i\beta\right)}
\left(\left(\lambda^2-\alpha^2-\beta^2\right)\mathbf{I}
+2\lambda\left( \begin{array}{cc} -i\Omega_1 & i\bar{u}
\\
 i u &  i\Omega_1  \\ \end{array} \right)\right)\right|_{x=0},
 \label{K3}
\end{eqnarray}
where $\alpha$ and $\beta$ are two real parameters, and
\begin{eqnarray}
\Omega_1=\sqrt{\beta^2 - |u|^2},
\label{Omg1}
\end{eqnarray}
we then obtain from (\ref{algc}) the following boundary condition
\begin{eqnarray}
\left.\left(iu_t-2u_{x}\Omega_1-(\alpha^2+\beta^2)u+2 u|u|^2\right)\right|_{x=0}=0.
\label{tbc}
\end{eqnarray}
In this case, the Poisson bracket (\ref{alga}) is equivalent to the following boundary Poisson bracket
\begin{eqnarray}
\left\{u,\bar{u}\right\}=-2i\Omega_1,
\label{BP2}
\end{eqnarray}
at $x=0$.
The boundary condition (\ref{tbc}) was derived in \cite{Zambon2014} via dressing a Dirichlet boundary with an integrable defect condition for the NLS equation (see e.g. \cite{CZ2006,Caudrelier2008,G2021}). The NLS equation in the presence of such a boundary condition attracted attention recently \cite{Xia2021,CCD2021,CCRZ2022,Gruner2020}.

Explicit forms of integrals of motion in involution and the Hamiltonian formulations for the above boundary models will be presented in section 3.3.

{\bf Remark} We emphasize that the boundary Poisson bracket (\ref{alga}) is essential for the Hamiltonian formulation of the boundary models and for the construction of the Poisson structure of the corresponding scattering data. This will become clear in section 3.3 and section 4.

\subsection{Inverse scattering transformation on the half-line}

The IST for the NLS equation in the presence of the aforementioned boundary conditions has been investigated recently in \cite{Zhangc2021} (see also \cite{CCRZ2022}). Here we will re-formulate this problem closely following the method and notations adopted in \cite{Faddeev2007} for full-line problem. Doing so is necessary, since it paves the way for studying the Poisson structure and Hamiltonian formulation for the half-line problem that will be performed in the next section, and it enables us to compare our results for the half-line problem with the case for full-line problem, more conveniently.
%This is necessary, since it will enable us in the next section to study the Poisson structures and Hamiltonian formulation for the half-line problem, more conveniently.
%Another improvement is that we treat the inverse part of the problem

\subsubsection{Jost solutions on the half-line}

Let
\begin{eqnarray}
F(x,y,\lambda)=T(x,y,\lambda)E(y,\lambda),
\label{Fxy}
\\
G(x,y,\lambda)=T(x,0,\lambda)K(\lambda)F(0,y,-\lambda)\tilde{K}(\lambda),
\label{Gxy}
\end{eqnarray}
where
\begin{eqnarray}
\tilde{K}(\lambda)=\diag\left(1,d(\lambda)\right),~~d(\lambda)=\det K(-\lambda).
\label{tK}
\end{eqnarray}

By using the integral representation (\ref{Tir}), we can prove that the limits
\begin{eqnarray}
F(x,\lambda)=\lim_{y\rightarrow\infty}F(x,y,\lambda),
\label{Fx}
\\
G(x,\lambda)=\lim_{y\rightarrow\infty}G(x,y,\lambda),
\label{Gx}
\end{eqnarray}
exist for real $\lambda$, and in particular $F(x,\lambda)$ has an integral representation
\begin{eqnarray}
F(x,\lambda)=E(x,\lambda)+\int_{x}^{\infty}\mathcal{F}(x,z)E(z,\lambda)dz,
\label{Fir}
\end{eqnarray}
where the kernel $\mathcal{F}(x,z)$ is given by
\begin{eqnarray}
\mathcal{F}(x,z)=-\frac{1}{2}U_0(\frac{x+z}{2})-\int_{\frac{x+z}{2}}^{\infty}\mathcal{T}^{\dag}(s,x,2s-z)U_0(s)ds,
\label{Fker}
\end{eqnarray}
and satisfies
\begin{eqnarray}
\int_{x}^{\infty}\|\mathcal{F}(x,z)\|dz
\leq \exp\left(\int_{x}^{\infty}\|U_0(z)\|dz\right)-1.
\label{Fkerest}
\end{eqnarray}
See Appendix A for details for the existence of the two limits and for the derivation of the integral representation (\ref{Fir}).

We now discuss the analytic properties of $F(x,\lambda)$ and $G(x,\lambda)$. The functions $F(x,y,\lambda)$ and $G(x,y,\lambda)$ are entire functions of $\lambda$. However, it is not the case for $F(x,\lambda)$ and $G(x,\lambda)$, since they involve a passage to the limit. Using the integral representation (\ref{Fir}), it follows that the first column of $F(x,\lambda)$ is bounded and analytic in the lower half of the complex $\lambda$-plane, while the second column of $F(x,\lambda)$ is bounded and analytic in the upper half of the complex $\lambda$-plane. We will denote the first and second columns of a $2\times 2$ matrix $A$ by $A^{(1)}$ and $A^{(2)}$, respectively.
We introduce the following notations
\begin{eqnarray}
\begin{split}
\Phi(x,\lambda)=F(x,\lambda)E(-x,\lambda),
\\
\Psi(x,\lambda)=G(x,\lambda)E(-x,\lambda).
\end{split}
\label{Phi}
\end{eqnarray}
The integral representation (\ref{Fir}) implies the following asymptotic behaviour for $\Phi(x,\lambda)$:
\begin{eqnarray}
\begin{split}
\Phi^{(1)}(x,\lambda)=&\left( \begin{array}{c} 1 \\ 0 \end{array}\right)+o(1), ~~\im\lambda \leq 0,~~|\lambda|\rightarrow\infty,
\\
\Phi^{(2)}(x,\lambda)=&\left( \begin{array}{c} 0 \\ 1 \end{array}\right)+o(1),~~\im\lambda \geq 0,~~|\lambda|\rightarrow\infty.
\end{split}
\label{Fasy}
\end{eqnarray}
Recall that $G(x,\lambda)$ involve the matrix $K(\lambda)$.
To avoid unnecessary singularities of $G(x,\lambda)$, we require that all zeros of $d(\lambda)$ are located in the lower half of the complex $\lambda$-plane. Then we can conclude that the first column of $\mathcal{G}\equiv K(\lambda)F(0,-\lambda)\tilde{K}(\lambda)$ is bounded and analytic in the upper half of the complex $\lambda$-plane, while the second column of $\mathcal{G}\equiv K(\lambda)F(0,-\lambda)\tilde{K}(\lambda)$ is bounded and analytic in the lower half of the complex $\lambda$-plane, and thus so does $G(x,\lambda)$. Moreover, we find from (\ref{Kasy}), (\ref{Tir}) and (\ref{Fir}) that $\Psi(x,\lambda)=G(x,\lambda)E(-x,\lambda)$ has the following asymptotic behaviour:
\begin{eqnarray}
\begin{split}
\Psi^{(1)}(x,\lambda)=&\left( \begin{array}{c} 1 \\ 0 \end{array}\right)+o(1), ~~\im\lambda \geq 0,~~|\lambda|\rightarrow\infty,
\\
\Psi^{(2)}(x,\lambda)=&\left( \begin{array}{c} 0 \\ 1 \end{array}\right)+o(1),~~\im\lambda \leq 0,~~|\lambda|\rightarrow\infty.
\end{split}
\label{Gasy}
\end{eqnarray}

For real $\lambda$, the functions $F(x,\lambda)$ and $G(x,\lambda)$ satisfy
\begin{eqnarray}
\det F(x,\lambda)=\det G(x,\lambda)=1,
\label{detFGx}
\\
F(x,\lambda)=\sigma \overline{F(x,\lambda)}\sigma,
~~G(x,\lambda)=\sigma \overline{G(x,\lambda)}\sigma.
\label{invFGx}
\end{eqnarray}
Indeed, the unimodular property $\det F(x,\lambda)=1$ follows from (\ref{detT}), the unimodular property $\det G(x,\lambda)=1$ follows from (\ref{detT}) and the normalization condition (\ref{dsym}). The involution relations (\ref{invFGx}) follow from the involution relation (\ref{Tinvol}) and the normalization conditions (\ref{Knorm}). The analytic properties of the columns of $F(x,\lambda)$ and $G(x,\lambda)$ imply that the involution property (\ref{invFGx}) may extend to complex values of $\lambda$, it takes the form
\begin{eqnarray}
\begin{split}
\overline{F^{(1)}(x,\lambda)}=i\sigma F^{(2)}(x,\bar{\lambda}),~~\im \lambda\leq 0,
\\
\overline{G^{(1)}(x,\lambda)}=i\sigma G^{(2)}(x,\bar{\lambda}),~~\im \lambda\geq 0.
\end{split}
\label{FGinv}
\end{eqnarray}

\subsubsection{Scattering data}

The functions $F(x,\lambda)$ and $G(x,\lambda)$ satisfy the differential equation (\ref{Tx}).
This in turn implies that they are related, namely
\begin{eqnarray}
G(x,\lambda)=F(x,\lambda)\Gamma(\lambda),
\label{GFx}
\end{eqnarray}
where $\Gamma(\lambda)$ is a matrix function independent of $x$. Evaluating (\ref{GFx}) at $x=0$, we obtain a representation of $\Gamma(\lambda)$,
\begin{eqnarray}
\begin{split}
\Gamma(\lambda)
=F_{0}^{-1}(\lambda)K(\lambda)F_{0}(-\lambda)\tilde{K}(\lambda),
\label{Gamx}
\end{split}
\end{eqnarray}
where
$F_{0}(\lambda)=\left.F(x,\lambda)\right|_{x=0}$.
The unimodular property (\ref{detFGx}) and the involution property (\ref{invFGx}) extend naturally to $\Gamma(\lambda)$,
\begin{eqnarray}
\det \Gamma(\lambda)=1,
\label{detGam}
\\
\Gamma(\lambda)=\sigma \overline{\Gamma(\lambda)}\sigma,
\label{invGam1}
\end{eqnarray}
for real $\lambda$.
In addition to the involution property (\ref{invGam1}), the matrix $\Gamma(\lambda)$ also satisfies
\begin{eqnarray}
\Gamma^{-1}(\lambda)=\tilde{K}(-\lambda) \Gamma(-\lambda)\tilde{K}(\lambda),
\label{invGam}
\end{eqnarray}
which can be verified by virtue of the normalization conditions (\ref{Knorm}) and (\ref{dsym}).
This symmetry relation is the main difference with respect to the full-line problem, and it is crucial to the analysis of the IST for the half-line problem.

The involution property (\ref{invGam1}) justifies the following notation for $\Gamma(\lambda)$,
\begin{eqnarray}
\Gamma(\lambda)=\left( \begin{array}{cc} A(\lambda) & -\overline{B(\lambda)} \\
 B(\lambda) & \overline{A(\lambda)}  \\ \end{array} \right),
\label{Gam}
\end{eqnarray}
for real $\lambda$.
It follows from (\ref{GFx}) that $A(\lambda)$ and $B(\lambda)$ can be expressed as
\begin{eqnarray}
\begin{split}
A(\lambda)=\det\left(G^{(1)}(x,\lambda),F^{(2)}(x,\lambda)\right),
\\
B(\lambda)=\det\left(F^{(1)}(x,\lambda),G^{(1)}(x,\lambda)\right),
\end{split}
\label{ABexp}
\end{eqnarray}
where as before $G^{(j)}(x,\lambda)$ and $F^{(j)}(x,\lambda)$, $j=1,2$ denote the columns of $G(x,\lambda)$ and $F(x,\lambda)$.
The analytic properties of $F^{(2)}(x,\lambda)$ and $G^{(1)}(x,\lambda)$ imply that $A(\lambda)$ has an analytic continuation into the upper half of the complex plane $\im\lambda\geq 0$.
The analytic properties of $F^{(1)}(x,\lambda)$ and $G^{(1)}(x,\lambda)$ imply that in general $B(\lambda)$ can only be well-defined for real $\lambda$.
The asymptotic formulae (\ref{Fasy}) and (\ref{Gasy}) imply the asymptotic behaviour
\begin{eqnarray}
\begin{split}
A(\lambda)=1+o(1),~~|\lambda|\rightarrow\infty,
\\
B(\lambda)=o(1),~~|\lambda|\rightarrow\infty.
\end{split}
\label{ABasymp}
\end{eqnarray}

With the notation (\ref{Gam}), equalities (\ref{detGam}) and (\ref{invGam}) become
\begin{eqnarray}
|A(\lambda)|^2+|B(\lambda)|^2 =1,~~\lambda\in \mathbb{R},
\label{ABunim1}
\\
A(\lambda)=\overline{A(-\lambda)},
~~B(\lambda)=-d(-\lambda)B(-\lambda), ~~\lambda\in \mathbb{R}.
\label{ABsym1}
\end{eqnarray}
We note that the first of (\ref{ABsym1}) can be analytically continued into the upper half of the complex $\lambda$-plane, that is
\begin{eqnarray}
A(\lambda)=\overline{A(-\bar{\lambda})}, ~~\im\lambda\geq 0.
\label{ABsym2}
\end{eqnarray}

To simplify our analysis we shall assume that $A(\lambda)$ has only finite number of simple zeros in the upper half of the complex $\lambda$-plane. The symmetry (\ref{ABsym2}) implies that the zeros of $A(\lambda)$ always appear in pairs: if $\lambda_j$ is a zero of $A(\lambda)$, then so does $-\overline{\lambda}_j$. We will exclude the special case that a zero of $A(\lambda)$ is a pure imaginary number for convenient sake. Let $\mathcal{Z}\equiv\left\{\lambda_j\right\}_{1}^{n}\cup\left\{-\overline{\lambda}_j\right\}_{1}^{n}$, $\re\lambda_j\neq 0$, $\im\lambda_j>0$, denote the set of the zeros of $A(\lambda)$. The expression (\ref{ABexp}) implies that for $\lambda=\lambda_j$, the first column of $G(x,\lambda)$ is proportional to the second column of $F(x,\lambda)$.
Let $\gamma_j$ be the proportionality coefficient,
\begin{eqnarray}
G^{(1)}(x,\lambda_j)=\gamma_jF^{(2)}(x,\lambda_j),~~j=1,\cdots,n.
\label{gam}
\end{eqnarray}
Let $\tilde{\gamma}_j$ be the proportionality coefficient corresponding to the paired zero $-\bar{\lambda}_j$ of $A(\lambda)$,
\begin{eqnarray}
G^{(1)}(x,-\bar{\lambda}_j)=\tilde{\gamma}_jF^{(2)}(x,-\bar{\lambda}_j),~~j=1,\cdots,n.
\label{gamtilde}
\end{eqnarray}
We can deduce that
\begin{eqnarray}
\gamma_j\bar{\tilde{\gamma}}_j=-d(-\lambda_j),~~j=1,\cdots,n.
\label{gamrel}
\end{eqnarray}
See appendix B for the proof of the relation.
It is clear that $\bar{\lambda}_j$, $-\lambda_j$, $1\leq j\leq n$, are the zeros of $A^*(\lambda)\equiv \overline{A(\bar{\lambda})}$ in the lower half-plane, and we have
\begin{eqnarray}
\begin{split}
G^{(2)}(x,\bar{\lambda}_j)=-\bar{\gamma}_jF^{(1)}(x,\bar{\lambda}_j),~~j=1,\cdots,n,
\\
G^{(2)}(x,-\lambda_j)=-\bar{\tilde{\gamma}}_jF^{(1)}(x,-\lambda_j),~~j=1,\cdots,n.
\end{split}
\label{gambar}
\end{eqnarray}

The quantity $A(\lambda)$ can be expressed in terms of its zeros and $B(\lambda)$ as
\begin{eqnarray}
A(\lambda)=\prod_{j=1}^N\frac{\left(\lambda-\lambda_j\right)\left(\lambda+\bar{\lambda}_j\right)}
{\left(\lambda-\bar{\lambda}_j\right)\left(\lambda+\lambda_j\right)}
\exp\left\{\frac{\lambda}{\pi i}\int_0^\infty\frac{\log\left(1-|B(\mu)|^2\right)}{\mu^2-\lambda^2}d\mu\right\},
~~\im\lambda>0.
\label{tracf}
\end{eqnarray}
To prove (\ref{tracf}), we consider the function
\begin{eqnarray}
\tilde{A}(\lambda)=A(\lambda)\prod_{j=1}^N\frac{\left(\lambda-\bar{\lambda}_j\right)\left(\lambda+\lambda_j\right)}
{\left(\lambda-\lambda_j\right)\left(\lambda+\bar{\lambda}_j\right)},
\label{Aexp}
\end{eqnarray}
which is analytic for $\im\lambda>0$ and has no zeros for $\im\lambda\geq0$.
It is easy to see
\begin{eqnarray}
|\tilde{A}(\lambda)|^2=|A(\lambda)|^2=1-|B(\lambda)|^2,
\label{ABtildeA}
\end{eqnarray}
for real $\lambda$.
Using (\ref{Aexp}) and (\ref{ABtildeA}) in the following formula (see e.g. \cite{Faddeev2007})
\begin{eqnarray}
\tilde{A}(\lambda)=\exp\left\{\frac{1}{\pi i}\int_{-\infty}^\infty\frac{\re\log\left(\tilde{A}(\mu)\right)}{\mu-\lambda}d\mu\right\}
\label{Atilde}
\end{eqnarray}
we obtain
\begin{eqnarray}
A(\lambda)=\prod_{j=1}^N\frac{\left(\lambda-\lambda_j\right)\left(\lambda+\bar{\lambda}_j\right)}
{\left(\lambda-\bar{\lambda}_j\right)\left(\lambda+\lambda_j\right)}
\exp\left\{\frac{1}{2\pi i}\int_{-\infty}^\infty\frac{\log\left(1-|B(\mu)|^2\right)}{\mu-\lambda}d\mu\right\}.
\label{tracf1}
\end{eqnarray}
This formula can be extended up to the real line using the Sochocki-Plemelj formula.
The equality (\ref{tracf}) follows from (\ref{tracf1}) after using the symmetry relation (\ref{ABsym1}) of $B(\lambda)$.

Following the terminology used in \cite{Faddeev2007} for the full-line problem, we will call $A(\lambda)$ and $B(\lambda)$ transition coefficients for the continuous spectrum, and will call $\gamma_j$, $\bar{\gamma}_j$, $j=1,\cdots,n$, transition coefficients for the discrete spectrum for the half-line problem. Note that $A(\lambda)$ and $B(\lambda)$ involve both the initial data and the boundary conditions, this can be seen from the expression (\ref{Gamx}) where $F_0(\lambda)$ encodes the initial data, while $K(\lambda)$ contains the boundary information.
The set $\{B(\lambda),\overline{B(\lambda)},\lambda_j,\bar{\lambda}_j,\gamma_j,\bar{\gamma}_j\}$ constitutes the so-called scattering data for our half-line problem.

\subsubsection{The time evolution of scattering data}

By using (\ref{algc}), (\ref{Td}) and the rapid decay of $u(x,t)$ as $x\rightarrow\infty$, we find that $\Gamma(\lambda)$ satisfies the evolution equation
\begin{eqnarray}
\frac{d\Gamma(\lambda)}{dt}=\frac{i\lambda^2}{2}\left[\sigma_3,\Gamma(\lambda)\right].
\label{Gamt}
\end{eqnarray}
In components, (\ref{Gamt}) can be written as
\begin{eqnarray}
\frac{d A(\lambda)}{dt}=0, ~~\frac{d B(\lambda)}{dt}=-i\lambda^2B(\lambda).
\label{ABt}
\end{eqnarray}
This in turn implies that the generating function for the conservation law is just $A(\lambda)$.
By studying large $\lambda$ expansion of $A(\lambda)$, we can extract explicit forms of the conserved quantities. This will be discussed in the next subsection.

Let us derive the time evolution of transition coefficients for the discrete spectrum.
It follows from (\ref{Td}) that
\begin{eqnarray}
\frac{d F(x,\lambda)}{dt}=V(x,\lambda)F(x,\lambda)-\frac{i\lambda^2}{2}F(x,\lambda)\sigma_3.
\label{Ft}
\end{eqnarray}
Using (\ref{algc}) and (\ref{Td}), we obtain
\begin{eqnarray}
\frac{d G(x,\lambda)}{dt}=V(x,\lambda)G(x,\lambda)-\frac{i\lambda^2}{2}G(x,\lambda)\sigma_3.
\label{Gt}
\end{eqnarray}
From (\ref{Ft}) and (\ref{Gt}), we have
\begin{eqnarray}
\frac{d F^{(2)}(x,\lambda_j)}{dt}=V(x,\lambda_j)F^{(2)}(x,\lambda_j)+\frac{i\lambda_j^2}{2}F^{(2)}(x,\lambda_j),
\label{F2t}
\\
\frac{d G^{(1)}(x,\lambda_j)}{dt}=V(x,\lambda_j)G^{(1)}(x,\lambda_j)-\frac{i\lambda_j^2}{2}G^{(1)}(x,\lambda_j).
\label{G1t}
\end{eqnarray}
The above two equations are compatible with (\ref{gam}) only if
\begin{eqnarray}
\frac{d \gamma_j}{dt}=-i\lambda_j^2\gamma_j,~~j=1,\cdots,n.
\label{gamjt}
\end{eqnarray}

In summary, we have formulated a transformation
\begin{eqnarray}
(u,\bar{u})\rightarrow (B(\lambda),\overline{B(\lambda)},\lambda_j,\bar{\lambda}_j,\gamma_j,\bar{\gamma}_j)
\label{transf}
\end{eqnarray}
from the functions $u$, $\bar{u}$ to the transition coefficients and discrete spectrum of the corresponding auxiliary linear problem.
In terms of the new variables, the NLS equation with the boundary conditions stated in section 3.1 can be easily solved:
\begin{eqnarray}
\begin{split}
B(\lambda,t)=\exp\left(-i\lambda^2(t-t_0)\right)B(\lambda,t_0),
\\
\lambda_j(t)=\lambda_j(t_0),~~\gamma_j(t)=\exp\left(-i\lambda_j^2(t-t_0)\right)\gamma_j(t_0),~~j=1,\cdots,n,
\end{split}
\label{Blamgamsol}
\end{eqnarray}
where $B(\lambda,t_0)$, $\lambda_j(t_0)$ and $\gamma_j(t_0)$ result from the initial data $u(x,t_0)$ by applying the transformation (\ref{transf}).
The inverse part of the problem can be solved by formulating an appropriate Riemann-Hilbert problem. We will not discuss this issue in this paper, since our main purpose is to investigate the Poisson structure and action-angle variables on the scattering data for the half-line problem. We refer the interested reader to \cite{Zhangc2021} and, in particular, to the monograph \cite{Fokas} for details regarding this issue.

{\bf Remark} For the class of exponentially fast decaying initial data, $B(\lambda)$ can be analytically off the real axis. In this case, $\gamma_j$ has a nice characterization: it can be expressed in terms of $B(\lambda)$ as $\gamma_j=B(\lambda_j)$ (this follows by substituting (\ref{gam}) into (\ref{ABexp})). We emphasize that this expression is not valid in general, since $B(\lambda)$ has no analytic continuation off the real line in the general case.

\subsection{Conserved quantities and trace identities}
Equations (\ref{ABt}) show that $A(\lambda)$ is independent on $t$. Thus, $A(\lambda)$ provides a generating function of the conserved quantities for the half-line problem with boundary conditions stated in section 3.1.
We now derive explicit forms of the conserved quantities by investigating the large $\lambda$ expansions of $\ln A(\lambda)$. For $|\lambda|\rightarrow \infty$, the transition matrix $T$ can be expanded as \cite{Faddeev2007}
\begin{eqnarray}
T(x,y,t,\lambda)=\left(\mathbf{I}+W(x,t,\lambda)\right)\exp\left[Z(x,y,t,\lambda)\right]
\left(\mathbf{I}+W(y,t,\lambda)\right)^{-1},
\label{texp}
\end{eqnarray}
where $W$ is an off-diagonal matrix and $Z$ is a diagonal matrix.
Inserting (\ref{texp}) into (\ref{lpx}), one may obtain that the elements of $W$ and $Z$ have the following asymptotic representations
\begin{eqnarray}
\begin{split}
W_{21}(x,t,\lambda)=\sum_{n=1}^{\infty}\frac{\omega_n(x,t)}{(i\lambda)^n}, ~~W_{12}(x,t,\lambda)=-\overline{W_{21}(x,t,\bar{\lambda})},
\\
Z_{22}(x,y,t,\lambda)=\frac{i}{2}\lambda(x-y)+\int_{y}^{x}u(z,t)W_{12}(z,t,\lambda)dz,
~~Z_{11}(x,y,t,\lambda)=\overline{Z_{22}(x,y,t,\bar{\lambda})},
\end{split}
\label{WZexp}
\end{eqnarray}
where the series coefficients $\omega_n$ are given recursively by
\begin{eqnarray}
\begin{split}
\omega_1=-u, ~~\omega_2=-u_x,
~~
\omega_{n+1}=\left(\omega_n\right)_x-\bar{u}\sum_{k=1}^{n-1}\omega_k\omega_{n-k}.
\end{split}
\label{omgn}
\end{eqnarray}

By calculating the $(11)$-element of (\ref{Gamx}), we obtain
\begin{eqnarray}
\ln A(\lambda)=\int_{0}^{\infty}\bar{u}(x,t)\left[W_{21}(x,t,-\lambda)-W_{21}(x,t,\lambda)\right]dx
+\ln\left(D_{11}(\lambda)\right),
\label{lnAexp}
\end{eqnarray}
where $D_{11}(\lambda)$ is the $(11)$-element of the matrix
\begin{eqnarray}
D(\lambda)=\left(\mathbf{I}+W(0,t,\lambda)\right)^{-1}K(\lambda)\left(\mathbf{I}+W(0,t,-\lambda)\right).
\label{D}
\end{eqnarray}
The first term in the right hand side of (\ref{lnAexp}) serves as a generating function for the bulk conserved quantities, while the second term provides a generating function for the boundary contribution to the conserved quantities. By computing the expansion of the logarithmic function $\ln\left(D_{11}(\lambda)\right)$ in $\left(i\lambda\right)^{-1}$,  %by virtue of the specific forms of the $K(\lambda)$ matrices presented in section
we may extract from (\ref{lnAexp}) explicit forms of the conserved quantities for the corresponding boundary problems order by order.
For example, the first two nontrivial conserved quantities for the NLS equation with the boundary condition (\ref{tbcnls}) are given by
\begin{eqnarray}
\begin{split}
I_1=2\int_{0}^{\infty}|u|^2dx+\left.\Omega\right|_{x=0},
\\
I_3=2\int_{0}^{\infty}\left(|u|^4-|u_x|^2\right)dx
+\left.\left(\frac{\Omega^3}{3}+2\Omega|u|^2\right)\right|_{x=0}.
\end{split}
\label{I131}
\end{eqnarray}
where $\Omega$ is defined by (\ref{Omg}).
The NLS equation with the boundary condition (\ref{tbcnls}) can be represented in the Hamilton form (\ref{HE}) by choosing the Hamiltonian to be
\begin{eqnarray}
H=-\frac{1}{2}I_3=-\int_{0}^{\infty}\left(|u|^4-|u_x|^2\right)dx
-\left.\left(\frac{\Omega^3}{6}+\Omega|u|^2\right)\right|_{x=0}.
\label{H1}
\end{eqnarray}
Indeed, using the Poisson bracket (\ref{cPB}) and the bulk part of Hamiltonian (\ref{H1}), we recover the NLS equation in bulk from (\ref{HE}), while using the boundary Poisson bracket (\ref{BP1}) and the boundary contribution to Hamiltonian (\ref{H1}) we recover the boundary condition (\ref{tbcnls}) from (\ref{HE}).
For the NLS equation with the boundary condition (\ref{tbc}), the first two nontrivial conserved quantities are
\begin{eqnarray}
\begin{split}
I_1=2\int_{0}^{\infty}|u|^2dx+2\left.\Omega_1\right|_{x=0},
\\
I_3=2\int_{0}^{\infty}\left(|u|^4-|u_x|^2\right)dx
+\left.\left(\frac{8(\Omega_1)^3}{3}+2\Omega_1\left(2|u|^2-\alpha^2-\beta^2\right)\right)\right|_{x=0}.
\end{split}
\label{I132}
\end{eqnarray}
where $\Omega$ is given by (\ref{Omg1}). The Hamiltonian associated with this boundary model can be recognized as
\begin{eqnarray}
H=-\frac{1}{2}I_3=-\int_{0}^{\infty}\left(|u|^4-|u_x|^2\right)dx
-\left.\left(\frac{4(\Omega_1)^3}{3}+\Omega_1\left(2|u|^2-\alpha^2-\beta^2\right)\right)\right|_{x=0}.
\label{H2}
\end{eqnarray}

We now show that the integrals of the motion can be represented in terms of the scattering data.
It follows from (\ref{tracf}) that the quantity $\ln A(\lambda)$ can be expanded as
\begin{eqnarray}
\ln A(\lambda)=\sum_{j=0}^{\infty}\frac{c_{2j+1}}{\left(i\lambda\right)^{2j+1}},
\label{Aexpansion1}
\end{eqnarray}
where
\begin{eqnarray}
c_{2j+1}=-\frac{1}{4^{j}\pi}\int_0^\infty\ln\left(1-|B(\mu)|^2\right)\left(2i\mu\right)^{2j}d\mu
+\frac{1}{4^{j}}\sum_{l=1}^{N}\frac{\left(2i\bar{\lambda}_l\right)^{2j+1}-\left(2i\lambda_l\right)^{2j+1}}{2j+1}.
\label{cj}
\end{eqnarray}
On the other hand, we write the expansion of (\ref{lnAexp}) as
\begin{eqnarray}
\ln A(\lambda)=\sum_{j=0}^{\infty}\frac{I_{2j+1}}{\left(i\lambda\right)^{2j+1}}.
\label{lnAexp2}
\end{eqnarray}
Comparing the asymptotic expansion (\ref{Aexpansion1}) with (\ref{lnAexp2}) yields the representations of the conserved quantities in terms of the scattering data
\begin{eqnarray}
I_{2j+1}=c_{2j+1},~~j=0,1,\cdots.
\label{Icj}
\end{eqnarray}
These identities are the trace identities for the half-line problem.
In particular, the Hamiltonian $H=-\frac{1}{2}I_3$ can be represented as
\begin{eqnarray}
H=-\frac{1}{2}c_3=-\frac{1}{2\pi}\int_0^\infty\mu^2\ln\left(1-|B(\mu)|^2\right)d\mu
+\frac{i}{3}\sum_{l=1}^{N}\left(\left(\bar{\lambda}_l\right)^{3}-\left(\lambda_l\right)^{3}\right).
\label{Hs}
\end{eqnarray}

The Poisson commutativity of the conserved quantities $I_{2j+1}$ generated from $\ln A(\lambda)$ will be proved in section 4 (see corollary 1 in section 4.1). Thus we can speak of the integrability of the boundary conditions stated in section 3.1 in the sense of the existence of infinitely many conserved quantities in involution.
%These conserved quantities will be shown in section 4 to be in involution (see (\ref{AABB}).

\section{Hamiltonian formulation for the half-line problem}

In this section, our aim will be to derive the Poisson brackets between all the elements
of the scattering data and to construct the variables of action-angle type for the half-line problem with boundary conditions stated in the above section.
%We note that in \cite{CCRZ2022} only the Poisson brackets for the continuous spectrum are computed. Here we give a full consideration: the Poisson brackets for the continuous spectrum, the Poisson brackets for the discrete spectrum and the Poisson brackets between the continuous and discrete spectrum are all derived. In particular, in order to derive the latter two Poisson brackets, the Poisson brackets between the Jost solutions (see proposition \ref{PBJ} below) are needed, whose derivation is much complicated than the one in the full-line case.

\subsection{Poisson structure on the scattering data}

Note that the $r$-matrix relation (\ref{rmrelation1}) can be extended to the case of transition matrices for two arbitrary intervals $(y,x)$ and $(y',x')$ (see e.g. \cite{Faddeev2007}), it reads
\begin{eqnarray}
\begin{split}
\left\{T_1(x,y,\lambda),T_2(x',y',\mu)\right\}=&\left(T(x,x'',\lambda)\otimes T(x',x'',\mu)\right)
\\&
\times\left[r(\lambda-\mu),T_1(x'',y'',\lambda)T_2(x'',y'',\mu)\right]
\\&
\times \left(T(y'',y,\lambda)\otimes T(y'',y',\mu)\right),
\end{split}
\label{grmrelation}
\end{eqnarray}
where $(y'',x'')$ means the intersection of the intervals $(y,x)$ and $(y',x')$.

\begin{lemma}
For $x<y$, we have the following Poisson bracket relations
\begin{eqnarray}
\begin{split}
\left\{F_1(x,y,\lambda),F_2(x,y,\mu)\right\}=&F_1(x,y,\lambda)F_2(x,y,\mu)r(y,\lambda-\mu)
-r(\lambda-\mu)F_1(x,y,\lambda)F_2(x,y,\mu),
\\
\left\{F_1(x,y,\lambda),G_2(x,y,\mu)\right\}=&F_1(x,y,\lambda)G_2(x,y,\mu)\tilde{r}(y,\lambda,\mu)
\\&-F_1(x,y,-\mu)G_2(x,y,\mu)\tilde{r}(\lambda,\mu)F^{-1}_1(x,y,-\mu)F_1(x,y,\lambda),
\\
\left\{G_1(x,y,\lambda),G_2(x,y,\mu)\right\}=&
r(\lambda-\mu)G_1(x,y,\lambda)G_2(x,y,\mu)-G_1(x,y,\lambda)G_2(x,y,\mu)\check{r}(y,\lambda,\mu)
\\&+G_1(x,y,\lambda)F_2(x,y,-\lambda)\hat{r}(\lambda,\mu)F^{-1}_2(x,y,-\lambda)G_2(x,y,\mu)
\\&-F_1(x,y,-\mu)G_2(x,y,\mu)\tilde{r}(\lambda,\mu)F^{-1}_1(x,y,-\mu)G_1(x,y,\lambda),
\end{split}
\label{FGyrmatrix}
\end{eqnarray}
where
\begin{eqnarray}
\tilde{r}(\lambda,\mu)=r(\lambda+\mu)\diag\left(1,d(\mu),d(-\mu),1\right),
\label{tr}
\\
\hat{r}(\lambda,\mu)=r(\lambda+\mu)\diag\left(1,d(-\lambda),d(\lambda),1\right),
\label{hr}
\\
r(y,\lambda-\mu)=\left(E(y,\mu-\lambda)\otimes E(y,\lambda-\mu)\right)r(\lambda-\mu),
\label{ry}
\\
\tilde{r}(y,\lambda,\mu)=\left(E(y,-\lambda-\mu)\otimes E(y,\lambda+\mu)\right)\tilde{r}(\lambda,\mu),
\label{tyr}
\\
\check{r}(y,\lambda,\mu)=\left(E(y,\lambda-\mu)\otimes E(y,\mu-\lambda)\right)r(\lambda-\mu)\diag\left(1,d(-\lambda)d(\mu),d(\lambda)d(-\mu),1\right).
\label{cry}
\end{eqnarray}
\end{lemma}
{\bf Proof} The first of (\ref{FGyrmatrix}) can be derived by using (\ref{rmrelation1}) or (\ref{rmrelation2}).
The second of (\ref{FGyrmatrix}) can be derived by using (\ref{grmrelation}).
The third of (\ref{FGyrmatrix}) can be derived by using (\ref{grmrelation}) together with the relation (\ref{alga}).
\QEDB

Remark that the terms in the right hand side of (\ref{FGyrmatrix}) involve $\frac{1}{\lambda\mp\mu}$ which is singular at $\lambda=\pm\mu$, so that we shall specify the generalized function $\frac{1}{\lambda\mp\mu}$ to be taken as p.v. $\frac{1}{\lambda\mp\mu}$, where p.v. indicates principal value, for definiteness. By taking the limits of (\ref{FGyrmatrix}) as $y\rightarrow\infty$ and by using the relation
\begin{eqnarray}
\lim_{y\rightarrow \infty} p.v.\frac{e^{\pm i\lambda y}}{\lambda}=\pm\pi i\delta(\lambda),
\end{eqnarray}
we obtain
\begin{proposition}\label{PBJ}
The Poisson brackets between the Jost solutions $F(x,\lambda)$ and $G(x,\lambda)$ are given by
\begin{eqnarray}
\begin{split}
\left\{F_1(x,\lambda),F_2(x,\mu)\right\}=&F_1(x,\lambda)F_2(x,\mu)r_{+}(\lambda-\mu)-r(\lambda-\mu)F_1(x,\lambda)F_2(x,\mu),
\\
\left\{F_1(x,\lambda),G_2(x,\mu)\right\}=&F_1(x,\lambda)G_2(x,\mu)\tilde{r}_{+}(\lambda,\mu)
-F_1(x,-\mu)G_2(x,\mu)\tilde{r}(\lambda,\mu)F^{-1}_1(x,-\mu)F_1(x,\lambda),
\\
\left\{G_1(x,\lambda),G_2(x,\mu)\right\}=&
r(\lambda-\mu)G_1(x,\lambda)G_2(x,\mu)-G_1(x,\lambda)G_2(x,\mu)r_{-}(\lambda-\mu)
\\&+G_1(x,\lambda)F_2(x,-\lambda)\hat{r}(\lambda,\mu)F^{-1}_2(x,-\lambda)G_2(x,\mu)
\\&-F_1(x,-\mu)G_2(x,\mu)\tilde{r}(\lambda,\mu)F^{-1}_1(x,-\mu)G_1(x,\lambda),
\end{split}
\label{FGrmatrix}
\end{eqnarray}
where
\begin{eqnarray}
\begin{split}
r_{+}(\lambda-\mu)=\lim_{y\rightarrow\infty}r(y,\lambda-\mu)
=\left( \begin{array}{cccc}
p.v.\frac{1}{\lambda-\mu} & 0 & 0 & 0
\\
0 &  0 & \pi i\delta(\lambda-\mu) & 0
\\
0 & -\pi i\delta(\lambda-\mu) & 0 & 0
\\
0 &  0 & 0 & p.v.\frac{1}{\lambda-\mu}
 \\ \end{array} \right),
\end{split}
\label{rm+}
\end{eqnarray}
$r_{-}(\lambda-\mu)$ differs from $r_{+}(\lambda-\mu)$ by replacing $i$ with $-i$, and
\begin{eqnarray}
\begin{split}
\tilde{r}_{+}(\lambda,\mu)=
\lim_{y\rightarrow\infty}\tilde{r}(y,\lambda,\mu)
=\left( \begin{array}{cccc}
p.v.\frac{1}{\lambda+\mu} & 0 & 0 & 0
\\
0 &  0 & \pi i\delta(\lambda+\mu)d(-\mu) & 0
\\
0 & -\pi i\delta(\lambda+\mu)d(\mu) & 0 & 0
\\
0 &  0 & 0 & p.v.\frac{1}{\lambda+\mu}
 \\ \end{array} \right),
\end{split}
\label{trm+}
\end{eqnarray}
and
$\tilde{r}(\lambda,\mu)$ and $\hat{r}(\lambda,\mu)$ are given by (\ref{tr}) and (\ref{hr}), respectively.
\end{proposition}

\begin{proposition}
For real $\lambda$ and $\mu$, we have the following Poisson bracket
relation
\begin{eqnarray}
\begin{split}
\left\{\Gamma_1(\lambda),\Gamma_2(\mu)\right\}=&
r_{+}(\lambda-\mu)\Gamma_1(\lambda)\Gamma_2(\mu)-\Gamma_1(\lambda)\Gamma_2(\mu)r_{-}(\lambda-\mu)
\\&-\Gamma_2(\mu)\tilde{r}_{+}(\lambda,\mu)\Gamma_1(\lambda)+\Gamma_1(\lambda)\hat{r}_{+}(\lambda,\mu)\Gamma_2(\mu),
\end{split}
\label{Gamrmatrix}
\end{eqnarray}
where $r_{+}(\lambda-\mu)$ is defined by (\ref{rm+}), $r_{-}(\lambda-\mu)$ differs from $r_{+}(\lambda-\mu)$ by replacing $i$ with $-i$, $\tilde{r}_{+}(\lambda,\mu)$ is defined by (\ref{trm+}), and $\hat{r}_{+}(\lambda,\mu)$ is  defined by
\begin{eqnarray}
\hat{r}_{+}(\lambda,\mu)
=&\left( \begin{array}{cccc}
p.v.\frac{1}{\lambda+\mu} & 0 & 0 & 0
\\
0 &  0 & -\pi i\delta(\lambda+\mu)d(\lambda) & 0
\\
0 & \pi i\delta(\lambda+\mu)d(-\lambda) & 0 & 0
\\
0 &  0 & 0 & p.v.\frac{1}{\lambda+\mu}
 \\ \end{array} \right).
\label{hrm+}
\end{eqnarray}
\end{proposition}
{\bf Proof} Straightforward calculations using (\ref{FGrmatrix}) give
\begin{eqnarray}
\begin{split}
&\left\{\Gamma_1(\lambda),\Gamma_2(\mu)\right\}
\\
=&r_{+}(\lambda-\mu)\Gamma_1(\lambda)\Gamma_2(\mu)-\Gamma_1(\lambda)\Gamma_2(\mu)r_{-}(\lambda-\mu)
\\
&-\Gamma_2(\mu)\tilde{r}_{+}(\lambda,\mu)\Gamma_1(\lambda)+\Gamma_1(\lambda)\hat{r}_{+}(\lambda,\mu)\Gamma_2(\mu)
\\&+\left(F_{0}^{-1}(\lambda)\otimes F_{0}^{-1}(\mu)\right)\hat{K}(\lambda,\mu)
\left(\left(F_{0}(-\lambda)\tilde{K}(\lambda)\right)\otimes \left(F_{0}(-\mu)\tilde{K}(\mu)\right)\right),
\end{split}
\label{Gamrmatrix1}
\end{eqnarray}
where
\begin{eqnarray*}
\hat{K}(\lambda,\mu)=\left\{K_1(\lambda),K_2(\mu)\right\}-\left[r(\lambda-\mu),K_1(\lambda)K_2(\mu)\right]
-K_1(\lambda)r(\lambda+\mu)K_2(\mu)+K_2(\mu)r(\lambda+\mu)K_1(\lambda).
\end{eqnarray*}
Inserting (\ref{alga}), that is $\hat{K}(\lambda,\mu)=0$, into (\ref{Gamrmatrix1}), we obtain (\ref{Gamrmatrix}).
\QEDB

\begin{proposition}\label{ABPoisson}
The matrix Poisson bracket (\ref{Gamrmatrix}) is equivalent to the following six basic ones
\begin{eqnarray}
\left\{A(\lambda),A(\mu)\right\}=0,~~\left\{A(\lambda),\overline{A(\mu)}\right\}=0,
\label{AABB}
\\
\left\{A(\lambda),B(\mu)\right\}=-\left(\frac{1}{\lambda-\mu+i0}+\frac{1}{\lambda+\mu+i0}\right)A(\lambda)B(\mu),
\label{AB}
\\
\left\{A(\lambda),\overline{B(\mu)}\right\}=\left(\frac{1}{\lambda-\mu+i0}+\frac{1}{\lambda+\mu+i0}\right)
A(\lambda)\overline{B(\mu)},
\label{ABbar}
\\
\left\{B(\lambda),B(\mu)\right\}=0,~~\left\{B(\lambda),\overline{B(\mu)}\right\}=2\pi i\left(\delta(\lambda-\mu)-\delta(\lambda+\mu)d(-\lambda)\right)|A(\lambda)|^2.
\label{BBbar}
\end{eqnarray}
\end{proposition}
{\bf Proof}
The proof can be completed by calculating explicit forms of the matrix Poisson bracket (\ref{Gamrmatrix}).
For example, calculating the $(21)$-entry of (\ref{Gamrmatrix}) after using the symmetry relations (\ref{ABsym1}),
we obtain
\begin{eqnarray}
\left\{A(\lambda),B(\mu)\right\}=\left(\pi i\delta(\lambda-\mu)-p.v.\frac{1}{\lambda-\mu}
+\pi i\delta(\lambda+\mu)-p.v.\frac{1}{\lambda+\mu}\right)A(\lambda)B(\mu),
\end{eqnarray}
which produces (\ref{AB}) by virtue of the Sochocki-Plemelj formula
\begin{eqnarray}
\frac{1}{\lambda\pm i0}=p.v.\frac{1}{\lambda}\mp \pi i\delta(\lambda).
\label{SPF}
\end{eqnarray}
Other Poisson bracket relations in this proposition can be verified via a similar manner.
\QEDB

From (\ref{AABB}), we immediately obtain
\begin{corollary}
The conserved quantities $I_{2j+1}$ constructed in section 3.3 are in involution,
\begin{eqnarray}
\left\{I_{2j+1},I_{2k+1}\right\}=0,~~j,k\geq 0.
\end{eqnarray}
\end{corollary}

We note that the relations (\ref{AABB}), (\ref{AB}) and (\ref{ABbar}) are consistent with the analyticity of $A(\lambda)$ in the upper half of the complex $\lambda$-plane, so that they can be analytically continued into the upper half of the complex $\lambda$-plane.

\begin{proposition}\label{ABgamj}
The Poisson brackets between the transition coefficients for the continuous spectrum and the ones for the discrete spectrum are given by
\begin{eqnarray}
\left\{B(\lambda),\gamma_j\right\}=\left\{B(\lambda),\bar{\gamma}_j\right\}=0,~~\lambda\in \mathbb{R}, ~~1\leq j\leq N,
\label{Bgam}
\\
\left\{A(\lambda),\gamma_j\right\}=-\frac{1}{\lambda+\lambda_j+i0}A(\lambda)\gamma_j
-\frac{1}{\lambda-\lambda_j}A(\lambda)\gamma_j,~~\im \lambda>0,
\label{Agam1}
\\
\left\{A(\lambda),\bar{\gamma}_j\right\}=\frac{1}{\lambda-\bar{\lambda}_j+i0}A(\lambda)\bar{\gamma}_j
+\frac{1}{\lambda+\bar{\lambda}_j}A(\lambda)\bar{\gamma}_j,~~\im \lambda>0,
\label{Agam2}
\end{eqnarray}
\end{proposition}
{\bf Proof}
We introduce the notation
\begin{eqnarray}
F^{(2)}(x,\lambda)=\left( \begin{array}{c}  f_{+}(x,\lambda) \\
 g_{+}(x,\lambda)  \\ \end{array} \right),
~~
G^{(1)}(x,\lambda)=\left( \begin{array}{c} f_{-}(x,\lambda) \\
 g_{-}(x,\lambda) \\ \end{array} \right).
\label{FG}
\end{eqnarray}
It follows from the unimodular property (\ref{detFGx}) that
\begin{eqnarray}
 |f_{\pm}(x,\lambda)|^2 + |g_{\pm}(x,\lambda)|^2=1,~~\lambda\in \mathbb{R}.
\label{FGunim1}
\end{eqnarray}
With the notation (\ref{FG}), the quantities $A(\lambda)$, $B(\lambda)$ and $\gamma_j$ can be expressed as
\begin{eqnarray}
\begin{split}
A(\lambda)=& g_{+}(x,\lambda)f_{-}(x,\lambda)-f_{+}(x,\lambda)g_{-}(x,\lambda),
\\
B(\lambda)=& f_{-}(x,\lambda)\overline{f_{+}(x,\lambda)}+g_{-}(x,\lambda)\overline{g_{+}(x,\lambda)},
\\
\gamma_j=&\left.\frac{f_{-}(x,\mu)}{f_{+}(x,\mu)}\right|_{\mu=\lambda_j}
=\left.\frac{g_{-}(x,\mu)}{g_{+}(x,\mu)}\right|_{\mu=\lambda_j}.
\end{split}
\label{ABGam}
\end{eqnarray}
After straightforward calculations using (\ref{FGrmatrix}) together with (\ref{ABsym1}) and (\ref{FGunim1}), we obtain that, for real $\lambda$ and $\mu$,
\begin{eqnarray}
\begin{split}
&\left\{B(\lambda),\frac{f_{-}(x,\mu)}{f_{+}(x,\mu)}\right\}
\\
=&
\frac{A(\mu)d(-\lambda)\overline{g_{+}(x,-\lambda)}f_{-}(x,-\lambda)}{f^2_{+}(x,\mu)}
\left(\pi i\delta(\lambda+\mu)-\frac{1}{\lambda+\mu}\right)
\\&
-\frac{A(\mu)f_{-}(x,\lambda)\overline{g_{+}(x,\lambda)}}{f^2_{+}(x,\mu)}
\left(\pi i\delta(\lambda-\mu)+\frac{1}{\lambda-\mu}\right).
\end{split}
\label{Bgp}
\end{eqnarray}
This equality can be analytically continued in $\mu$, so that we may set $\mu=\lambda_j$.
By doing so, the term in the left hand side of (\ref{Bgp}) becomes $\left\{B(\lambda),\gamma_j\right\}$, while the terms in the right hand side of (\ref{Bgp}) vanishes due to $A(\lambda_j)=0$. Hence, $\left\{B(\lambda),\gamma_j\right\}=0$.
Proceeding as above, we obtain
\begin{eqnarray}
\begin{split}
\left\{A(\lambda),\gamma_j\right\}=\left(\pi i\delta(\lambda+\lambda_j)-\frac{1}{\lambda+\lambda_j}\right)A(\lambda)\gamma_j
-\frac{1}{\lambda-\lambda_j}A(\lambda)\gamma_j,
\end{split}
\end{eqnarray}
which becomes (\ref{Agam1}) according to the Sochocki-Plemelj formula (\ref{SPF}).
The remaining Poisson brackets in this proposition can be derived via a similar manner.
\QEDB

Regarding the rest of the Poisson brackets between the scattering data, we have
\begin{proposition}\label{disPoisson}
The following Poisson bracket relations hold
\begin{eqnarray}
\left\{\gamma_j,\gamma_k\right\}=\left\{\gamma_j,\bar{\gamma}_k\right\}=0,~~1\leq j,k\leq N,
\label{gamjk}
\\
\left\{\lambda_j,\lambda_k\right\}=\left\{\lambda_j,\bar{\lambda}_k\right\}=0,~~1\leq j,k\leq N,
\label{lamjk}
\\
\left\{B(\mu),\lambda_j\right\}=\left\{B(\mu),\bar{\lambda}_j\right\}=0,~~1\leq j\leq N,
\label{Blamjk}
\\
\left\{\gamma_j,\bar{\lambda}_k\right\}=0,~~\left\{\lambda_k,\gamma_j\right\}=\delta_{jk}\gamma_j,~~1\leq j,k\leq N.
\label{gamlam}
\end{eqnarray}
\end{proposition}
{\bf Proof} The Poisson brackets (\ref{gamjk}) can be derived by using the Poisson brackets (\ref{FGrmatrix}) and the expression of $\gamma_j$ given by (\ref{ABGam}). The Poisson brackets (\ref{lamjk}) follow from (\ref{AABB}).
We rewrite (\ref{AB}) as
\begin{eqnarray}
\left\{\log A(\lambda),B(\mu)\right\}=-\left(\frac{1}{\lambda-\mu}+\frac{1}{\lambda+\mu}\right)B(\mu),
\label{AB1}
\end{eqnarray}
where it is assumed that $\im\lambda>0$.
Inserting (\ref{Aexp}) into (\ref{AB1}), we obtain
\begin{eqnarray}
\begin{split}
&\left\{\log\tilde{A}(\lambda),B(\mu)\right\}
+\sum_{j=1}^N\left(\frac{\left\{B(\mu),\lambda_j\right\}}{\lambda-\lambda_j}
+\frac{\left\{B(\mu),\lambda_j\right\}}{\lambda+\lambda_j}
+\frac{\left\{\bar{\lambda}_j,B(\mu)\right\}}{\lambda-\bar{\lambda}_j}
+\frac{\left\{\bar{\lambda}_j,B(\mu)\right\}}{\lambda+\bar{\lambda}_j}
\right)
\\=&-\frac{B(\mu)}{\lambda-\mu}-\frac{B(\mu)}{\lambda+\mu}.
\end{split}
\label{AB2}
\end{eqnarray}
The right hand side of equation (\ref{AB2}) is analytic for $\im \lambda>0$, thus the left hand side has no singularities at $\lambda=\lambda_j$ and at $\lambda=-\bar{\lambda}_j$. Therefore, relations (\ref{Blamjk}) hold.
Proceeding as above, we obtain, after using (\ref{Agam1}) and (\ref{Aexp}), the following equation
\begin{eqnarray}
\begin{split}
\left\{\log\tilde{A}(\lambda),\gamma_j\right\}
+\sum_{k=1}^N\left(\frac{\left\{\gamma_j,\lambda_k\right\}}{\lambda-\lambda_k}
+\frac{\left\{\gamma_j,\lambda_k\right\}}{\lambda+\lambda_k}
+\frac{\left\{\bar{\lambda}_k,\gamma_j\right\}}{\lambda-\bar{\lambda}_k}
+\frac{\left\{\bar{\lambda}_k,\gamma_j\right\}}{\lambda+\bar{\lambda}_k}
\right)
=-\frac{\gamma_j}{\lambda-\lambda_j}-\frac{\gamma_j}{\lambda+\lambda_j}.
\end{split}
\label{Agam3}
\end{eqnarray}
For $\im \lambda>0$, the right hand side of equation (\ref{Agam3}) has singularities at $\lambda=\lambda_j$.
Comparing the residues at $\lambda=\lambda_j$, we find the second of relations (\ref{gamlam}).
The left hand side of (\ref{Agam3}) should have no singularities at $\lambda=-\bar{\lambda}_j$ (since the right hand side has no singularities at $\lambda=-\bar{\lambda}_j$), this fact implies the first of relations (\ref{gamlam}).
\QEDB

It follows from propositions \ref{ABPoisson}, \ref{ABgamj} and \ref{disPoisson} that the non-vanishing Poisson brackets of the scattering data $\{B(\lambda),\overline{B(\lambda)};\lambda_j,\bar{\lambda}_j,\gamma_j,\bar{\gamma}_j;j=1,2,\cdots,N\}$
are
\begin{eqnarray}
\left\{B(\lambda),\overline{B(\mu)}\right\}=2\pi i\left(\delta(\lambda-\mu)-\delta(\lambda+\mu)d(-\lambda)\right)\left(1-|B(\lambda)|^2\right),
\label{BBbarnon}
\\
\left\{\lambda_k,\gamma_j\right\}=\delta_{jk}\gamma_j,~~1\leq j,k\leq N.
\label{lg}
\end{eqnarray}
We note that the Poisson bracket (\ref{BBbarnon}) is quite different from the one for the full-line problem.
Indeed, in the full-line case the non-vanishing Poisson bracket for the transition coefficient $b(\lambda)$ reads \cite{Faddeev2007}
\begin{eqnarray}
\left\{b(\lambda),\overline{b(\mu)}\right\}=2\pi i\left(1-|b(\lambda)|^2\right)\delta(\lambda-\mu).
\label{bbp}
\end{eqnarray}
Compared to (\ref{bbp}), the Poisson bracket (\ref{BBbarnon}) involves both $\delta(\lambda-\mu)$ and $\delta(\lambda+\mu)$, and particularly it involves $d(\lambda)$ which encodes the boundary information for the half-line problem.

\subsection{Variables of action-angle type}

Let us introduce the quantities
\begin{eqnarray}
\begin{split}
\rho\left(\lambda\right)=-\frac{1}{2\pi}\ln\left(1-|B(\lambda)|^2\right),
~~\phi\left(\lambda\right)=-\arg B(\lambda),~~\lambda\in \mathbb{R},
\\
p_j=2\re \lambda_j,~~q_j=\log|\gamma_j|,~~1\leq j,k\leq N,
\\
\varrho_j=2\im \lambda_j,~~\varphi_j=-\arg\gamma_j,~~1\leq j,k\leq N.
\end{split}
\label{cvariables}
\end{eqnarray}
After straightforward calculations using (\ref{ABsym1}) and the Poisson brackets in propositions \ref{ABPoisson}, \ref{ABgamj} and \ref{disPoisson}, we find that the Poisson brackets between the above quantities have the form
\begin{eqnarray}
\begin{split}
\left\{\rho\left(\lambda\right),\phi\left(\mu\right)\right\}=\delta(\lambda-\mu)+\delta(\lambda+\mu),
~~\left\{\rho\left(\lambda\right),\rho\left(\mu\right)\right\}
=\left\{\phi\left(\lambda\right),\phi\left(\mu\right)\right\}=0,
\\
\left\{p_j,q_k\right\}=\delta_{jk},~~\left\{p_j,p_k\right\}=\left\{q_j,q_k\right\}=0,~~1\leq j,k\leq N,
\\
\left\{\varrho_j,\varphi_k\right\}=\delta_{jk},~~\left\{\varrho_j,\varrho_k\right\}=\left\{\varphi_j,\varphi_k\right\}=0,
~~1\leq j,k\leq N.
\end{split}
\label{cPoisson}
\end{eqnarray}
The quantities (\ref{cvariables}) constitute the action-angle variables for the NLS equation on the half-line.
The main distinction from the full-line case is that the first of Poisson brackets (\ref{cPoisson}) involves both $\delta(\lambda-\mu)$ and $\delta(\lambda+\mu)$.

Expressions (\ref{cj}) and (\ref{Icj}) give the conserved quantities in terms of the action variables  $\rho\left(\lambda\right)$, $p_j$ and $\varrho_j$:
\begin{eqnarray}
I_{2j+1}=2(-1)^j\int_0^\infty\rho(\mu)\mu^{2j}d\mu
+\frac{1}{4^{j}\left(2j+1\right)}\sum_{l=1}^{N}\left[\left(ip_l+\varrho_l\right)^{2j+1}-\left(ip_l-\varrho_l\right)^{2j+1}\right].
\label{Ijnew}
\end{eqnarray}
In particular, the Hamiltonian $H=-\frac{1}{2}I_3$ can be expressed in terms of the action variables as
\begin{eqnarray}
H=\int_0^\infty\mu^2\rho\left(\mu\right)d\mu
+\frac{1}{12}\sum_{l=1}^{N}\varrho_l\left(3p^2_l-\varrho_l^2\right).
\label{Hs1}
\end{eqnarray}

From (\ref{cPoisson}) and (\ref{Hs1}), we immediately have
\begin{proposition}
The action-angle variables (\ref{cvariables}) completely trivialize the dynamics of the NLS equation on the half-line:
\begin{eqnarray}
\begin{split}
\frac{d\rho\left(\lambda\right)}{dt}=\left\{H,\rho\right\}=0,
~~\frac{d p_j}{dt}=\left\{H,p_j\right\}=0,
~~\frac{d \varrho_j}{dt}=\left\{H,\varrho_j\right\}=0,
\\
\frac{d\phi\left(\lambda\right)}{dt}=\left\{H,\phi\right\}=\lambda^2,
~~\frac{d q_j}{dt}=\left\{H,q_j\right\}=\frac{1}{2}\varrho_jp_j,
\\
\frac{d \varphi_j}{dt}=\left\{H,\varphi_j\right\}=\frac{1}{4}\left(p^2_j-\varrho^2_j\right),
~~j=1,\cdots, N.
\end{split}
\label{HE1}
\end{eqnarray}
\end{proposition}
We note that the above formulae agree with the time evolutions of the scattering data already established in section 3 (see (\ref{ABt}) and (\ref{gamjt})).

\section{Concluding remarks}

We derive the action-angle variables for the NLS equation on the half-line subjecting to a class of integrable boundary conditions. These variables are expressed in terms of the scattering data for this integrable half-line model. A related and important question is the study of Poisson structures and action-angle variables for integrable differential-difference equations, such as the Ablowitz-Ladik lattice system, on the set of non-negative integers. We will investigate this issue in the near future.

\section*{ACKNOWLEDGMENTS}
This work was supported by the National Natural Science Foundation of China (Grant No. 12271221).

\begin{appendices}

\section{Existence of limits (\ref{Fx}) and (\ref{Gx}) and a derivation for the integral representation (\ref{Fir})}
Using (\ref{Tir}), we find that $F^{-1}(x,y,\lambda)$ has the integral representation
\begin{eqnarray}
F^{-1}(x,y,\lambda)=E(-x,\lambda)+\int_{x}^{2y-x}E(-z,\lambda)\mathcal{T}(y,x,z)dz,
\label{Fyir}
\end{eqnarray}
where the kernel $\mathcal{T}(y,x,z)$ satisfies
\begin{eqnarray}
\mathcal{T}(y,x,z)=\frac{1}{2}U_0(\frac{x+z}{2})+\int_{\frac{x+z}{2}}^{y}U_0(s)\mathcal{T}(s,x,2s-z)ds.
\label{Fyker}
\end{eqnarray}
Let $\tilde{\mathcal{T}}(y,x)=\int_{x}^{2y-x}\|\mathcal{T}(y,x,z)\|dz$.
Then we obtain, after using (\ref{Fyker}) and interchanging the integrals, the estimate
\begin{eqnarray}
\tilde{\mathcal{T}}(y,x)\leq \int_{x}^{y}\|U_0(z)\|dz+\int_{x}^{y}\|U_0(s)\|\tilde{\mathcal{T}}(s,x)ds.
\label{Test}
\end{eqnarray}
By iterating the above estimate, we obtain
\begin{eqnarray}
\tilde{\mathcal{T}}(y,x)\leq \exp\left(\int_{x}^{y}\|U_0(z)\|dz\right)-1.
\label{Test1}
\end{eqnarray}
We claim that the limit
\begin{eqnarray}
\mathcal{T}_{+}(x,z)=\lim_{y\rightarrow\infty}\mathcal{T}(y,x,z)
\end{eqnarray}
defines a function of $z$ in $L_{1}^{2\times 2}(x,\infty)$ for each fixed $x$. Indeed, by using (\ref{Fyker}) and (\ref{Test1}), we find that
\begin{eqnarray}
\begin{split}
\int_{x}^{\infty}\|\mathcal{T}_{+}(x,z)\|dz
\leq &\int_{x}^{\infty}\|U_0(z)\|dz+\int_{x}^{\infty}\|U_0(s)\|\tilde{\mathcal{T}}(s,x)ds
\\
\leq &\exp\left(\int_{x}^{\infty}\|U_0(z)\|dz\right)-1.
\end{split}
\label{kerest}
\end{eqnarray}
The above analysis implies that the limit (\ref{Fyir}), as $y\rightarrow\infty$, does exist, and there is an integral representation
\begin{eqnarray}
F^{-1}(x,\lambda)=E(-x,\lambda)+\int_{x}^{\infty}E(-z,\lambda)\mathcal{T}_{+}(x,z)dz,
\label{Firinv}
\end{eqnarray}
where the kernel $\mathcal{T}_{+}(x,z)$ is given by
\begin{eqnarray}
\mathcal{T}_{+}(x,z)=\frac{1}{2}U_0(\frac{x+z}{2})+\int_{\frac{x+z}{2}}^{\infty}U_0(s)\mathcal{T}(s,x,2s-z)ds.
\label{Fkerinv}
\end{eqnarray}
Due to $F(x,\lambda)$ is unimodular, we have $F^{-1}(x,\lambda)=\sigma F^{\tau}(x,\lambda)\sigma$, where the superscript $\tau$ denotes the transpose of a matrix. Using this relation, we find the integral representation (\ref{Fir}) for $F(x,\lambda)$.
%where the kernel $\mathcal{F}(x,z)$ is given by
%\begin{eqnarray}
%\mathcal{F}(x,z)=-\frac{1}{2}U_0(\frac{x+z}{2})-\int_{\frac{x+z}{2}}^{\infty}\mathcal{T}^{\dag}(s,x,2s-z)U_0(s)ds.
%\label{Fker}
%\end{eqnarray}
The existence of the limit (\ref{Fx}) implies the existence of the limit $\lim_{y\rightarrow\infty}F(0,y,-\lambda)$.
Thus the limit (\ref{Gx}) also exists.

\section{A proof for the relation (\ref{gamrel})}
We may set $x=0$ in (\ref{gam}) and (\ref{gamtilde}) to characterize $\gamma_j$ and $\tilde{\gamma}_j$, since these two quantities are independent on $x$.
With the notation (\ref{FG}), we have
\begin{eqnarray}
\gamma_j\bar{\tilde{\gamma}}_j=\frac{f_{-}(0,\lambda_j)}{f_{+}(0,\lambda_j)}
\frac{\overline{f_{-}(0,-\bar{\lambda}_j)}}{\overline{f_{+}(0,-\bar{\lambda}_j)}}.
\label{gamrel1}
\end{eqnarray}
By the definition of $G(x,\lambda)$, we have
\begin{eqnarray}
G(0,\lambda)=K(\lambda)F(0,-\lambda)\tilde{K}(\lambda).
\label{G0}
\end{eqnarray}
With the notation (\ref{FG}), we obtain from (\ref{G0}) the following expressions
\begin{eqnarray}
\begin{split}
\overline{f_{-}(0,-\bar{\lambda})}
=& d(-\lambda)\left(K_{21}(-\lambda)f_{+}(0,\lambda) +K_{22}(-\lambda)g_{+}(0,\lambda)\right),
\\
\overline{f_{+}(0,-\bar{\lambda})}
=& \frac{1}{d(-\lambda)}\left(K_{21}(\lambda)f_{-}(0,\lambda) -K_{11}(\lambda)g_{-}(0,\lambda)\right),
\end{split}
\label{f0}
\end{eqnarray}
where $K_{jk}(\lambda)$, $j,k=1,2$, stand for the $jk$-entries of the matrix $K(\lambda)$.
The normalization condition (\ref{Knorm}) gives $K(-\lambda)=K^{-1}(\lambda)$, which yields
\begin{eqnarray}
K_{21}(\lambda)=-d(-\lambda)K_{21}(-\lambda),~~K_{22}(\lambda)=d(-\lambda)K_{11}(-\lambda).
\label{Ksym}
\end{eqnarray}
Inserting (\ref{f0}) and (\ref{Ksym}) into (\ref{gamrel1}), we obtain
\begin{eqnarray}
\gamma_j\bar{\tilde{\gamma}}_j=-d(-\lambda_j)
\frac{K_{21}(\lambda_j)f_{-}(0,\lambda_j)f_{+}(0,\lambda_j)-K_{11}(\lambda_j)g_{+}(0,\lambda_j)f_{-}(0,\lambda_j)}
{K_{21}(\lambda_j)f_{-}(0,\lambda_j)f_{+}(0,\lambda_j)-K_{11}(\lambda_j)g_{-}(0,\lambda_j)f_{+}(0,\lambda_j)}
=-d(-\lambda_j),
\label{gamrel2}
\end{eqnarray}
where, in the last equality, we have used $g_{+}(0,\lambda_j)f_{-}(0,\lambda_j)=g_{-}(0,\lambda_j)f_{+}(0,\lambda_j)$.
This completes the proof for the relation (\ref{gamrel}).

\end{appendices}

\vspace{1cm}
\small{

}
\end{document}